\documentclass{aa}
\usepackage{graphicx}
\usepackage{amsmath}
\usepackage{txfonts}
\usepackage{rotating}
\begin{document}
   \title{Searching for star-forming dwarf galaxies \\ in the Antlia cluster 
          \thanks{Based on observations acquired at Gemini South (GS-2010A-Q-51 and GS-2012A-Q-59) and ESO VISTA Hemisphere Survey (VHS)}
         }

      \author{O. Vaduvescu~\inst{1}
          \and
              C. Kehrig~\inst{2}
          \and
              L.~P. Bassino~\inst{3,4,5}
          \and
              A. V. Smith Castelli~\inst{3,4,5}
          \and
              J.~P. Calder\'on~\inst{3,4,5}
              }

   \offprints{O. Vaduvescu}

   \institute{Isaac Newton Group of Telescopes, Apto. 321, E-38700 Santa Cruz de la Palma, Canary Islands, Spain\\
              \email{ovidiuv@ing.iac.es}
            \and
              Instituto de Astrof\'isica de Andaluc\'ia (CSIC), Apto. 3004, 18080, Granada, Spain\\
            \and
              Grupo de Investigaci\'on CGGE, Facultad de Ciencias Astron\'omicas y Geof\'isicas, Universidad Nacional de La Plata, Paseo del Bosque, B1900FWA La Plata, Argentina\\
	    \and
              Consejo Nacional de Investigaciones Cient\'ificas y T\'ecnicas (CONICET), C1033AAJ, Ciudad Aut\'onoma de Buenos Aires, Argentina\\
            \and
              Instituto de Astrof\'isica de La Plata (CCT-La Plata, CONICET-UNLP), Paseo del Bosque, B1900FWA La Plata, Argentina\\
             }

   \date{Accepted for publication in A\&A (early 2014)}

 
  \abstract
  { 
   The formation and evolution of dwarf galaxies in clusters need to be understood,
   and this requires large aperture telescopes. 
  } 
  { In this sense, we selected the Antlia cluster to continue our previous work in the Virgo, Fornax, 
    and Hydra clusters and in the Local Volume (LV). Because of the scarce available literature data, 
    we selected a small sample of five blue compact dwarf (BCD) candidates in Antlia for observation. } 
  { Using the Gemini South and GMOS camera, we acquired the $H\alpha$ imaging needed to detect 
    star-forming regions in this sample. With the long-slit spectroscopic data of the brightest seven 
    knots detected in three BCD candidates, we derived their basic chemical properties. Using archival 
    VISTA VHS survey images, we derived $K_S$ magnitudes and surface brightness profile fits for the 
    whole sample to assess basic physical properties. } 
  { FS90-98, FS90-106, and FS90-147 are confirmed as BCDs and cluster members, based on their morphology, 
   $K_S$ surface photometry, oxygen abundance, and velocity redshift. FS90-155 and FS90-319 did not show any 
   $H\alpha$ emission, and they could not be confirmed as dwarf cluster star-forming galaxies. Based on 
   our data, we studied some fundamental relations to compare star forming dwarfs (BCDs and dIs) in the LV 
   and in the Virgo, Fornax, Hydra, and Antlia clusters. 
  } 
  { Star-forming dwarfs in nearby clusters appear to follow same fundamental relations in the 
    near infrared with similar objects in the LV, specifically the size-luminosity and the 
    metallicity-luminosity, while other more fundamental relations could not be checked in Antlia due 
    to lack of data. 
   } 

   \keywords{galaxies -- dwarf, blue compact dwarf, irregular, formation, evolution, 
             fundamental parameters, photometry, structure; infrared -- galaxies}

   \authorrunning{Vaduvescu et al.}
   \maketitle

%

\section{Introduction}

Blue compact dwarf galaxies (BCD) are part of the vast family of dwarf that 
also includes dwarf elliptical (dE), dwarf irregular (dI), and dwarf 
spheroidal (dSph) galaxies. BCDs show quite distinctive characteristics such 
as intense star formation at their central regions and low metallicity (e.g., 
Kehrig et al. \cite{keh08}; Cairos et al. \cite{cai09}), but most of 
them also harbor an old stellar population that accounts for most of their 
mass (Papaderos et al. \cite{pap96}; Cair\'os et al. \cite{cai01,cai02}; Noeske 
et al. \cite{noe03,noe05}; Vaduvescu et al. \cite{vad06}; Amor\'in et al. \cite{amo09}). 
Some BCDs carry important information related to the first stages of 
star formation (Sargent \& Searle \cite{sar70}; Lequeux \&  Viallefond \cite{leq80}; 
Kunth \& Sargent \cite{kun86}) and the study of nearby BCDs located in the local 
universe should be used first to elucidate properties of high-redshift, low-mass galaxies 
that are harder to reach. 

According to the observational criteria proposed by Thuan \& Martin (\cite{thu81}) 
and Gil de Paz, Madore \& Pevunova (\cite{gil03}), BCDs have low luminosity 
($M_B \gtrsim -18$ mag and $M_K \gtrsim -21$ mag) and low peak surface brightness 
($\mu_{B} < 22~\rm{mag~arcsec^{-2}}$), and their integrated colours are mainly 
blue (mean $B-R \approx 0.7$ mag). Spectroscopically, they show intense and narrow 
emission lines super-imposed on a faint and blue optical continuum (Kehrig et al. 
\cite{keh04}). The oxygen abundance takes values, for instance, in the range 
$ 7.49 \leq 12 + \rm{log(O/H)} \leq 8.81$ dex in a sample of about 70 nearby 
field BCDs selected from the Gil de Paz et al. (\cite{gil03}) sample by Zhao, Gao 
\& Gu (\cite{zha10}), though abundances were obtained using different methods (see 
also Cairos et al. \cite{cai10}; Kehrig et al. \cite{keh13}). These results support the 
statement that they are metal-poor systems, since their oxygen abundances are mostly 
below the solar value ($12 + \rm{log(O/H)} \approx 8.7~$ dex (Asplund et al. \cite{asp09}). 

A correlation between the oxygen abundance and the absolute $K_S$ magnitude seems to hold 
for BCDs and dIs both in the Local Volume (LV), defined as the nearby Universe within $d<11$ Mpc 
(Karachentsev et al. \cite{kar13}), and in the nearby clusters (Vaduvescu et al. \cite{vad07,vad11}). 
Furthermore, the near infrared (NIR) luminosity of the old 
stellar population in star-forming dwarfs can be used to estimate the gas fraction 
(Vaduvescu et al. \cite{vad05,vad06}), defined as the gas mass over total baryonic 
mass (i.e., gas and stars). Vaduvescu et al. (\cite{vad11}) confirmed that the 
abundance - gas fraction relation for LV BCDs seems to be followed by star-forming dwarfs 
located in Virgo, Fornax, and Hydra clusters. 
Based on these results, they propose that the chemical evolution of BCDs in the 
Virgo and Hydra clusters seems to be consistent with the predictions of a closed-box 
model. They also suggest that the mass-metallicity relation followed by star-forming 
dwarfs looks different for environments of different densities. The dwarfs in clusters 
follow a mass-metallicity relation with a slope steeper than the one of the same 
relation for LV dwarfs. More recently, Zhao, Gao \& Gu (\cite{zha13}) have examined 
the oxygen abundance-gas fraction relation obtained for a sample of 53 BCDs and 22 field 
dIs and compared it with different models of chemical evolution. They conclude that most 
galaxies in their sample do not agree with a closed-box model; i.e., BCDs do not seem 
to have evolved as isolated systems. Since the dispersion in most of these relations is 
rather large, it is essential to increase the sample in order to determine the influence 
of the environment. 

During the past decades, a few possible evolutionary connections between different 
types of dwarfs  (BCDs, dIs, dEs, dSphs) have been explored, taking into account 
that several physically different channels are at work, but the whole evolution 
scenario between them remains far from clearly understood (e.g., Kunth 
\& \"{O}stlin \cite{kun00}; Vaduvescu et al. \cite{vad11} and references therein; 
S\'anchez-Janssen et al. \cite{san13}). 
This paper follows Vaduvescu, McCall \& Richer (\cite{vad06}) and 
Vaduvescu, Richer, and McCall (\cite{vad07}) who focussed on Virgo cluster, and 
Vaduvescu et al. (\cite{vad11}), who focussed on the Fornax and Hydra clusters, all 
with the aim of studying star-forming dwarfs (BCDs and dIs) in nearby clusters and 
comparing their evolution with a field sample (Vaduvescu et al. \cite{vad05}; 
McCall et al. \cite{mcc12}). 
Following this investigation, in this sense we now scrutinize one small sample 
of star-forming dwarf candidate galaxies located in the southern cluster Antlia. 
The paper is organized as follows: Section~2 describes the cluster and object 
selection, and Section~3 presents the observations and data reduction. The results
are presented in Section~4 and discussed in Section~5. Finally, Section~6 summarizes 
the main conclusions derived from this work. 


\section{Cluster and target selection}

To address possible environmental effects of dwarf formation and evolution, 
star-forming dwarfs located in nearby clusters should be studied first, and the most 
accessible clusters remain the closest ones ($d<100$ Mpc). 

The Antlia cluster (or Abell S636) is the third nearest well populated cluster 
($d=35.16 \pm 1.78$ Mpc - Dirsch et al. \cite{dir03}) to our Local Group, 
with only the Virgo cluster (minimum distance $d\ge16.5$ Mpc; Gavazzi et al. 
\cite{gav03,gav05}) and Fornax ($d=20.0 \pm 1.7$ Mpc, Blakeslee et al. \cite{bla09}), 
and closer than Hydra ($d=46 \pm 5$ Mpc, Jensen at al. \cite{jen99}). 
Antlia is a moderately compact group classified as a rare Bautz-Morgan type III cluster, 
meaning that the cluster has no dominating brightest galaxy. Located in the Hydra-Centaurus 
supercluster, Antlia shows an elongated morphology with two major concentrations dominated 
by the massive and equally bright elliptical galaxies NGC 3268 and NGC 3258. 

Using spectral data, Hopp and Materne (\cite{hop85}) measured radial velocities of 258 
galaxies in the direction of Antlia cluster within a large region spanning $10\time10$ 
square degrees. This region sample was based on the Lauberts (\cite{lau82})) photographic 
catalogue built using plates taken with the ESO 1-m Schmidt telescope in La Silla. This catalogue 
consists mostly of spiral ($\sim$\,$75\%$) and elliptical galaxies ($\sim$\,$15\%$), with about 
$10\%$ classified as irregular and including $\sim$\,$5\%$ dwarfs. 
The Antlia cluster itself was defined dynamically by Hopp and Materne (\cite{hop85}) by 
fitting a probability density function to the observed Lauberts galaxy distribution. Besides 
the main group (named Antlia I), these authors identified four small galaxy systems (``mini-clusters'') 
from which the Antlia II system alone could compare in total luminosity and mass with Antlia I. 
The mean radial velocity of early type and late type galaxies of the Antlia I system are similar, 
which is different from the Virgo cluster. For all five mini-clusters the authors derived mean 
velocities, dispersions, total luminosities, virial masses, virial radii, and mass-to-light 
ratios (see Table V of Hopp and Materne, \cite{hop85}). 

The first photographic survey of Antlia was published by Ferguson \& Sandage (\cite{fer90}, 
abbreviated FS90 catalogue) who identified 375 galaxies by visual inspection in a search 
considered complete to $B_T$\,$\sim20$ mag using plates taken with the 2.5m duPont telescope in
Las Campanas Observatory. 
A membership status and a morphological type was assigned to each galaxy of the FS90 catalogue 
(1: definite member, 2: likely member, 3: possible member). From the total of 375 galaxies, 
232 are classified as definite or likely members. Forty-nine galaxies are classified as irregular 
($13\%$) and only seven are classified as BCDs ($2\%$). The density profile of the cluster 
(counted in total number of galaxies per square degree) drops off steeply with radius, this 
trend being similar with those of Fornax, Virgo, and other clusters (Ferguson and Sandage, 
\cite{fer90} - see their Figures 15 and 16). 

Nakazawa et al. (\cite{nak00}) studied the ASCA and ROSAT satellite observations of Antlia, 
concluding that the cluster is bright in the X-ray and almost isothermal within $50^\prime$ 
diameter around NGC~3268 (i.e., one of the two peak emissions visible on both ASCA and ROSAT 
X-ray images), deriving an average cluster temperature of $kT \sim 2.0$ keV. 
The modelled metal abundance is $Z \sim 0.35 Z_\odot$, and the cluster luminosity is 
$3.4 \time 10^{42} h_{75}^{-2}$ erg $s^{-1}$ within $r \le 250 h_{75}^{-1}$ kpc
(Nakazawa et al. \cite{nak00}). There is no X-ray central excess brightness and little 
central cool component, from which one could conclude that the potential profile of Antlia 
is simple, because it lacks the hierarchical structure often seen in other nearby poor clusters 
dominated by one single central galaxy.

During the past decade, the study of Antlia has been continued with CCD photometric 
and spectroscopic studies of the galaxies' stellar content, part of the Argentinian-Chilean 
Antlia Cluster Project based mainly on data obtained with the 4m Blanco telescope at CTIO, 
the VLT, and Gemini 8m telescopes, as well as imaging data taken from the HST archive. 
This Antlia Cluster Project covers not only star-forming galaxies, but also globular clusters 
(Dirsch et al. \cite{dir03}, Bassino et al. \cite{bas08}), ultra-compact dwarfs (UCDs: Caso 
et al. \cite{cas13}; Caso et al., in preparation) and galaxy populations (Smith Castelli et 
al. \cite{smi08a,smi08b,smi12}; Calder\'on et al., in preparation; Bassino et al., 
in preparation). 

Based on the literature mentioned above (specifically the FS90 catalogue), 
for the present work we selected the best five BCD Antlia cluster candidates, 
namely FS90-98, FS90-106, FS90-147, FS90-155, and FS90-319. 
All of them were classified as ``BCD'' or ``BCD?''  in the FS90 catalogue and were selected 
because they show star-forming knots on the MOSAIC Antlia Project images obtained with MOSAIC 4-m Blanco 
telescope (CTIO). Though FS90-98 and FS90-106 were originally classified by FS90 as ``status 3'' 
(i.e., possible cluster member), they were later spectroscopically confirmed to be members of Antlia 
(Smith Castelli et al. \cite{smi08a}). The rest of the objects, FS90-147, FS90-155, and FS90-319, 
are candidates with membership ``status 2'' (i.e., likely member). The two confirmed BCDs share 
the same locus at the colour-magnitude diagram of FS90 galaxies in the cluster core, on the 
blue side of the colour-magnitude relation defined by the early-type Antlia galaxies (Smith 
Castelli et al. \cite{smi08a}, see their Figure 2), where the ``blue cloud'' of star-forming 
galaxies is expected (e.g., Bell et al \cite{bel07}). 

In this paper we adopt a distance modulus for the Antlia cluster $(m - M) = 32.73 \pm 0.25$ 
(Dirsch et al. \cite{dir03}) determined based on the surface brightness fluctuation (SBF) and 
fundamental plane methods for the two large elliptical members NGC~3258 and NGC~3268, which 
corresponds to the cluster distance $d = 35.16$ Mpc, so that 1 arcsec subtends 170 pc.


\section{Observations}
\label{observations}

Two proposals were submitted by our team in 2009 and 2011 and observed by the Gemini South 
8.1m telescope in service mode in 2010A (Run ID: GS-2010A-Q-51) and 2012A (Run ID: GS-2012A-Q-59) 
in both imaging and spectroscopy modes. Additionally, VISTA Hemisphere Survey (VHS) $K_S$ band 
images from the ESO Image Archive\footnote{http://archive.eso.org/eso/eso\_archive\_main.html} were used. 

\subsection{Gemini $H\alpha$ \& $R$ Imaging}

During the first Gemini run, we used the Gemini Multi-Object Spectrograph camera (GMOS-S; 
Hook et al. \cite{hoo04}) to acquire $H\alpha$ imaging of the sample. 
The GMOS-S camera consists of three $2048 \times 4608$ pixel CCDs with pixel size 
$13.5~\mu$m, resulting in a $6144 \times 4608$ pixel mosaic covering $5.5\arcmin \times 
5.5\arcmin$ on-sky field. We used $2 \times 2$ binning (pixel scale of $0 \farcs 146$) 
that yielded a total field of view of $10\farcm0 \times 7\farcm5$. 
All observations were taken in poor weather Band 3 conditions defined 
as image quality $85\%$ (Zenith seeing $R\le1.05^{\prime\prime}$), clouds cover $70\%$ 
(patchy clouds or extended thin cirrus), sky background $80\%$ (grey) and any water vapour. 

For all targets, we used the Ha\_G0336 and r\_G0326 Gemini filters (briefly, $H\alpha$ 
and $R$, respectively). The observed seeing in $R$ was between $0.8^{\prime\prime}-1.0^{\prime\prime}$. 
The observing log of our Gemini run is given in Table~\ref{table1}. 

We processed the $H\alpha$ and $R$ pre-imaging frames in IRAF\footnote{IRAF is distributed 
by the National Optical Astronomical Observatories, which are operated by the Association 
of Universities for Research in Astronomy, Inc., under cooperative agreement with 
the National Science Foundation.} (only the central CCD including our small 
targets) using reduced bias and flat field images, then we combined the individual frames 
to produce the final reduced images. The ``net'' $H\alpha$ images were continuum-subtracted 
by measuring about 20 stars in each field, whose magnitudes were scaled in order 
to subtract $R$ images from the original $H\alpha$ images. For the $R$ images we derived 
zero points using 10-15 stars in each field with available $R$ magnitudes from the UCAC4 
catalogue that were expected to have magnitude errors smaller than 0.1 mag (Zacharias et al. 
\cite{zac12}). Our determined zero points agree with published Gemini night zero points 
within 0.1 mag, thus UCAC4 magnitudes are worthy for achieving this photometric precision. 
We present the reduced Gemini $R$, $H\alpha$ and pure $H\alpha$ images in Figure~\ref{fig1} 
(left, middle, and right panels, respectively). 

\subsection{VISTA $K_S$ Imaging}

The $K_S$ imaging covering Antlia cluster was observed on 5 Feb 2013 by the VHS. 
VISTA 4.1m telescope uses the VIRCAM camera, which consists of 16 inter-spaced Raytheon 
VIRGO HgCdTe infrared detectors $2048\times2048$ pixels having a pixel scale of 
$0.34\arcsec$ and covering each $11.6\arcmin\times11.6\arcmin$ field of view. The 
seeing was $1.0\arcsec$. 

We queried the ESO Image Archive, downloaded and reduced VISTA raw $K_S$ images covering 
each galaxy, resulting in four images for each target exposed for 15s, thus 1min total exposure 
time for each galaxy. The VISTA VSH survey remains quite shallow for dwarf galaxies, nevertheless 
sufficient to detect in $K_S$ our five targets and derive their magnitudes and $K_S$ surface 
brightness profiles. 

NIR image reduction was performed in IRAF. The sky was subtracted using the 
raw image closest in time to any raw science image, in order to remove the rapid sky 
and instrument signature variations. To reduce the $K_S$ photometry, we calculated 
zero points based on 10-15 available 2MASS stars in each of the VISTA CCD fields, 
resulting in zero point errors less than 0.1 mag for each reduced field. 
We present the reduced VISTA $K_S$ images in Figure~\ref{fig2}. 

\subsection{Gemini spectroscopy}

During the second Gemini run, follow-up spectroscopic data were obtained for the 
best three BCD candidates, using the same GMOS-S camera. The observed galaxies are 
listed in Table~\ref{table1}. 
We used the B600 grating centred at $4610$~\AA ~in the blue and the R400 grating 
centred at $6120$~\AA ~to cover the red side of the spectrum. The data were binned 
by a factor of 2 in the spatial and the spectral dimensions, yielding a scale of 
$0.146~^{\prime\prime}$/pixel. The spectral resolution is $5.0$~\AA~FWHM and 
$7.0$~\AA~FWHM on the blue and the red sides, 
respectively. We took the spectra of the galaxies using a slit width of $1.0\arcsec$. 
Observations of the spectrophotometric standard star LTT~3864 were obtained during 
the observed nights for flux calibration. Bias frames, dome flat-fields and CuAr arc 
exposures were taken as part of the Gemini baseline calibrations. 

We reduced the GMOS long-slit spectra using the standard Gemini IRAF routines. More 
details on the GMOS data reduction procedure can be found in Vaduvescu et al. (\cite{vad11}). 
We extracted the 1D spectra from the wavelength-calibrated and cosmic-ray-subtracted 2D 
spectra. The flux calibration of the 1D spectra was performed using the observations of 
the flux standard star LTT~3864. 
The flux-calibrated spectra for our galaxies are displayed in Figure~\ref{fig3}. 


\section{Results}
\label{results}

\subsection{Imaging and surface photometry}

\subsubsection{Gemini $R$ and $H\alpha$ imaging}

We present the Gemini $R$ images in the left-hand panel of Figure~\ref{fig1} (field of 
view FOV $1^\prime \times 1^\prime$ and normal sky orientation). The middle panel 
includes the reduced $H\alpha$ images (FOV $0.5^\prime \times 0.5^\prime$), and 
the net $H\alpha$ images are included in the right-hand panel (FOV $0.5^\prime \times 
0.5^\prime$). 

Only three galaxies show clear net $H\alpha$ emission, namely FS90-98, FS90-106, and 
FS90-147. Consequently, we selected them as best BCD candidates and have acquired 
Gemini spectroscopy for them for the remaining proposed time. Although the other two 
objects, FS90-155 and FS90-319, produced clear details in both $R$ and $H\alpha$ images  
(a few brighter knots and some fainter structure), both failed to show any clear net 
measurable $H\alpha$ emission. This is a possible consequence of either 
the poor observing conditions (causing imperfect continuum removal), doubt about cluster 
membership (background/foreground galaxies), or the non-starburst nature of these galaxies. 

We studied the surface photometry of the Gemini $R$ images, employing the \textsc{STSDAS} 
package (\textsc{ELLIPSE} task) in IRAF fixing the three parameters (centre, ellipticity, 
and position angle) in an iterative approach. The surface brightness profiles in $R$ are 
plotted in the right-hand panel of Figure~\ref{fig2} together with error bars (visible only 
on the outskirts). Total apparent magnitudes in $R$ were measured by integrating 
growing ellipses and are listed in Table~\ref{table2}. 

\subsubsection{VISTA $K_S$ imaging}

The reduced VISTA $K_S$ images are presented in the left-hand panel of Figure~\ref{fig2} 
(FOV $1^\prime \times 1^\prime$, matching the $R$ field, and normal sky orientation). 
All five objects were detected by VISTA in 1 min, with FS90-106 and FS90-147 being 
the faintest objects. 

Because of the much shallower VISTA $K_S$ images compared with Gemini $R$ ones, we 
matched the NIR ellipse fitting parameters (center, ellipticity, and position angle) with 
the deeper ones in $R$, in order to perform the shallower $K_S$ surface photometry 
using more solid initial conditions. NIR bands and particularly $K_S$ are better than 
visible bands in tracing galaxy stellar mass, because the old star population tracing 
most of the mass is mostly visible in the NIR, while young stars dominate the flux in 
the visible (e.g., Vaduvescu et al. \cite{vad05}). 

Despite the short integration time of VISTA imaging, in Section~\ref{discussion} we 
employ the NIR fits to study surface photometry in order to characterize galaxy profiles 
and classify them. Following our previous BCD work, we modelled all NIR profiles with two 
components, namely a ``sech'' function (hyperbolic secant) to account for the old extended 
component discovered to fit dIs well (Vaduvescu et al \cite{vad05}) plus a Gaussian to account 
for the outburst close to center in a BCD (Vaduvescu, Richer \& McCall \cite{vad06}). To 
derive the two components, we used the \textsc{NFIT1D} function of the \textsc{FITTING} 
package under IRAF. 

Given its importance {\bf for our study}, we call the definition of the ``sech'' fit 
(Vaduvescu et al \cite{vad05}): 

\begin{equation}
\label{law_flux}
I = I_0 \hbox{ sech} {(r/r_0)} = \frac{I_0}{ \cosh (r/r_0)} = \frac{2I_0}{e^{r/r_0}+e^{-r/r_0}}
\end{equation}

\noindent
Here $I$ represents the fitted flux at radius $r$ (defined as the distance from the centre along 
the semi-major axis), $I_0$ is the {\it central intensity} (expressed in counts/pixel), and $r_0$ 
represents the {\it scale length} of the profile (expressed in pixels). One should note that both 
these two ``sech'' fitting parameters have the same meaning as the central intensity and scale 
radius fit by the classical exponential, de Vaucouleur, or S\'ersic models. In magnitude units, 
the sech fit is given by the following equation: 
\begin{equation}
\label{law_mag}
\begin{split}
\mu = zp_s - 2.5 \log(I) = zp_s - 2.5 \log{(I_0\hbox{ sech}(r/r_0))} \\
  = zp_s - 2.5 \log{\frac{I_0}{\cosh(r/r_0)}}
\end{split}
\end{equation} 

\noindent
where $zp_s$ represents the zero point of the surface brightness magnitude system. 
At larger radii, equation (\ref{law_mag}) is convergent to the {\it sech magnitude} ($m_S$), while 
curving and levelling out at near-zero radii toward the {\it central surface brightness} ($\mu_0$). 
Expressed in mag arcsec$^{-2}$, the central surface brightness is simply
\begin{equation}
\label{law_magzero} 
\mu_0 = zp_s - 2.5 \log(I_0)
\end{equation}

In the right-hand panel of Figure~\ref{fig2} we plot the $K_S$ sech fit, the Gaussian, 
and the total of the two components which closely follows the profiles of BCDs from the 
outskirts to the centre. 
Error bars in $K_S$ are larger than those in $R$, and surface brightness profiles sample 
about three times less than their $R$ deeper Gemini counterparts. 

Our estimated errors are less than $0.3^{\prime\prime}$ in position centre, about 0.1 
in ellipticity, and less than ten degrees in position angle. Within these errors, based on 
an independent analysis of the $R$ profiles using variable fitting parameters made by one
of us (JPC), the ellipse fitting parameters in $K_S$ and $R$ bands agree very well, 
despite the much shallower $K_S$ profiles. 

\subsubsection{Physical parameters}

In Table~\ref{table2} we include the physical parameters derived in $K_S$ and $R$. 
Ellipticity $e$ and position angle $PA$ (measured from the north positively 
counter-clockwise) are given in the columns 2 and 3. Total apparent magnitudes $m_{TK}$ 
in $K_S$ (column 4) and $m_{TR}$ in $R$ (column 9) were measured by integrating growing 
ellipse apertures, and $R-K_S$ total colours (column 10) were derived from them. 
Sech magnitudes $m_{SK}$ (column 5) were derived by modeling the $K_S$ profiles with the 
sech + Gaussian, where $\mu_{0K}$ (column 6) represents the sech central surface brightness 
expressed in mag/arcsec$^2$ and $r_{0K}$ (column 7) is the sech scale radius expressed in 
arcsec. The calculated sech semi-major radius $r_{22K}$ (column 8, in arcsec) at $m_K=22$ 
mag/arcsec$^2$ are listed next. The sech absolute magnitudes in $K_S$, M$_{SK}$ 
(column 11) were calculated from sech $m_{SK}$ using the adopted distance modulus 
for the cluster $(m - M) = 32.73$ mag. The hypothetical absolute magnitudes of the last 
two galaxies were calculated under the assumption of cluster membership (neither confirmed 
based on the available data nor on our undetected $H\alpha$ flux). 
We plotted the colour profiles (up to the shallower $K_S$ limit) in the right-hand panel 
of Figure~\ref{fig2}. 

\subsubsection{Comparison with the literature}

Few data are available in the literature to compare with our derived total magnitudes. 
Using Blanco MOSAIC imaging in Kron-Cousins $R$ band and $1^{\prime\prime}$ seeing, 
Smith Castelli et al (\cite{smi08a}) derived for FS90-98 $R = 15.92$ mag, which comes 
very close to our result ($R = 16.02$ mag), while for FS90-106 same authors derived 
$R = 17.08$ mag, again very similar to our measurements ($R = 17.02$ mag). 
Two objects were detected by 2MASS, namely FS90-155 whose 2MASS $K_S = 13.26$ magnitude 
coincides with our VISTA measurement, and the extremely faint galaxy FS90-319 whose 
shallower 2MASS measurement ($K_S = 13.76$ mag) comes quite close to our VISTA 
measurements ($R = 13.16$ mag). 

\subsection{Spectroscopic analysis}

\subsubsection{Cluster membership}

Averaging our measurements obtained from the two bright lines, $5007~[OIII]$ and 
$6563~H\alpha$, we obtain the following redshift and velocities for the three galaxies 
observed in spectroscopy mode: for FS90-98 $z = 0.0095,~v = 2850$ km/s, 
for FS90-106 $z = 0.0086,~v = 2580$ km/s, and for FS90-147 $z = 0.0062,~v = 1860$ km/s. 
Compared to the cluster average velocity and $3~\sigma$ dispersions of the sub-clusters 
detected by Hopp and Materne (\cite{hop85} - see their Table 5), we can confirm cluster 
membership for all three galaxies. 

\subsubsection{Line fluxes}

For each spectrum we measured fluxes and equivalent widths (EW) of the emission 
lines using the Gaussian profile-fitting option in the IRAF task \textsc{SPLOT}. 
The intensity of each emission line was derived by integrating between two points 
given by the position of a local continuum. The line-flux errors were calculated 
using the following expression (e.g., Castellanos \cite{cas00}; Kehrig et al. \cite{keh08}): 

\begin{equation} 
\sigma_{line}=\sigma_{cont}N^{1/2}\left(1 + \frac{\rm EW}{N\Delta\lambda}\right)^{1/2}
\end{equation}

\noindent
where $\sigma_{cont}$ is the standard deviation of the continuum near the emission 
line, $N$ is  the width of the region used to measure the line in pixels, $\Delta\lambda$ 
is the spectral  dispersion in \AA/pixel, and EW represents the equivalent width of the line. 

We computed the reddening coefficient, C(H$\beta$), from the ratio of the 
measured-to-theoretical H$\alpha$/H$\beta$, while simultaneously solving for the effects 
of underlying Balmer absorption and assuming case B recombination for T$_{e}$ = 10$^{4}$ K 
and n$_{e}$ = 100 cm$^{-3}$ (Storey \& Hummer 1995). Formal errors in the C$(H\beta)$ 
were computed by propagating uncertainties in line fluxes, $I(H\alpha)$ and $I(H\beta)$: 

\begin{equation}
\sigma_{c(H \beta)}=\sigma_{I_{H\alpha}/I_{H\beta}}
\left[\frac{1}{ln(10)[f(H\alpha)-f(H\beta)](I_{H\alpha}/I_{H\beta})}\right]
\end{equation} 

Table~\ref{table3} lists the de reddened emission line fluxes relative to H$\beta$ for the 
observed objects along with their extinction coefficient C(H$\beta$). 

\subsubsection{Nebular analysis}
\label{nebular}

We obtained the electron densities, n$_{e}$, from the [SII]$\lambda\lambda$6717/6731 
line ratio following the five-level atom FIVEL software (Shaw \& Dufour, \cite{sd94}) available 
in the IRAF task \textsc{TEMDEN}. The derived n$_{e}$ values place all galaxies in the 
low-density regime (n$_{e}$ $\le$ 100 cm$^{-3}$). 

To obtain direct oxygen abundance measurements, one should first derive the electron 
temperature, which requires the measurement of faint auroral lines. We could not detect any 
temperature-sensitive line in our spectra (such as [OIII]$\lambda$4363). 
In this case, the alternative is to use strong line-abundance indicators (e.g., Pagel \cite{pag79}; 
Kehrig et al. \cite{keh06}; Pilyugin et al. \cite{pvt10}; Marino et al. \cite{mar13}).
To estimate the oxygen abundance we applied the widely used N2 parameter 
[N2 $\equiv$ log ([NII]$\lambda$6584/H$\alpha$)] following the empirical calibration proposed 
by Marino et al. (\cite{mar13}). This parameter does not depend on reddening effects or flux 
calibration issues owing the close wavelength of the two lines involved. In addition, the 
relationship between N2 and O/H is single-valued. Recently, Marino et al. (\cite{mar13}) 
have revised the N2 index using a new dataset comprising 603 extragalactic HII regions that 
represents the most comprehensive compilation of Te-based abundances in external galaxies to date. 
The uncertainty in the metallicity determination based on this calibration amounts to $\sim$ 0.2 dex. 
Oxygen abundances for our sample galaxies are given in Table~\ref{table3}. 


\section{Discussion}
\label{discussion}

In this section we discuss our findings obtained from the Gemini and VISTA imaging 
data (Figures~\ref{fig1} and \ref{fig2}) and spectroscopy data (Table~\ref{table3}).
We compare our results with other (very scarce) literature findings. 

\subsection{Morphological analysis}
\label{morph}

FS90-98 resembles to an iE galaxy, defined by Loose \& Thuan (\cite{loo86}) as 
a compact galaxy showing several knots closed to centre superposed on a larger low 
surface brightness non-star-forming elliptical component. 
It shows two non-central star-forming knots (noted 1 and 2) separated by 
$4.4^{\prime\prime}$ in the net $H\alpha$ right-hand panel image of Figure~\ref{fig1}. 
The colour profile appears slightly bluer towards the outskirts, with total colour of 
$R - K_S = 2.02$ mag. The $K_S$ profile could be fitted with a sech plus a Gaussian 
component, consistent with the BCD classification of this galaxy. 

FS90-106 shows a similar iE appearance in both $R$ and $H\alpha$ images 
with at least two non-central knots visible in the $H\alpha$ image separated 
by $2.5^{\prime\prime}$. The total color $R - K_S = 1.12$ is very blue, consistent 
with a BCD classification. Nevertheless, the $K_S$ image is the faintest in the 
whole sample, so the $K_S$ apparent magnitude and colour profile should be 
regarded with caution outside $\sim3^{\prime\prime}$ radius. 

FS90-147 clearly shows a few non-central knots in the Gemini $R$ and $H\alpha$ 
images, labelled along the two slits A and B in Figure~\ref{fig1}. The most visible 
are the four central bright knots in the $H\alpha$ image. The brighter pair 
(along slit A) is separated by $2.9\arcsec$, while the fainter pair along slit B 
is separated by $4.2\arcsec$ (nodes marked as B1 and B2). Other diffuse regions 
are visible to the north and south in the $H\alpha$ image, from which node 
B3 was also placed along slit B. 
The $K_S$ image is quite shallow and shows extended diffuse NIR emission,
while the $R - K_S$ profile seems to become redder towards the outskirts sampled 
areas, and total colour is $R - K_S = 2.25$ mag. All these findings, together with 
the surface profile in $K_S$ make the BCD classification probable for this galaxy. 

FS90-155 clearly shows a nucleus close to centre in the $R$, $H\alpha$ and 
$K_S$ images, having no $H\alpha$ emission. Its margins appear well defined 
towards the north-east, while towards the south this galaxy shows some flocculences 
(more visible visible in the original FITS image), which resembles more of a spiral galaxy 
with counter-clockwise winding arms. The colour profile varies between 
$1.5 < R - K_S < 3.0$ with total colour $R - K_S = 2.43$ mag. 

FS90-319 shows two brighter knots superposed in the North of an elliptical diffuse 
emission in all $R$, $H\alpha$ and $K_S$ images, although no $H\alpha$ emission 
could be detected. Due to the brighter knot in the North-East, both $R$ and $K_S$ 
profiles show a peak around $6^{\prime\prime}$ radius, and the colour profile appears 
flat at $R - K_S \sim 2.4$ mag.

\subsection{Chemical abundance analysis}

\begin{figure}[p]
\centering
\includegraphics[angle=0,width=13.6cm]{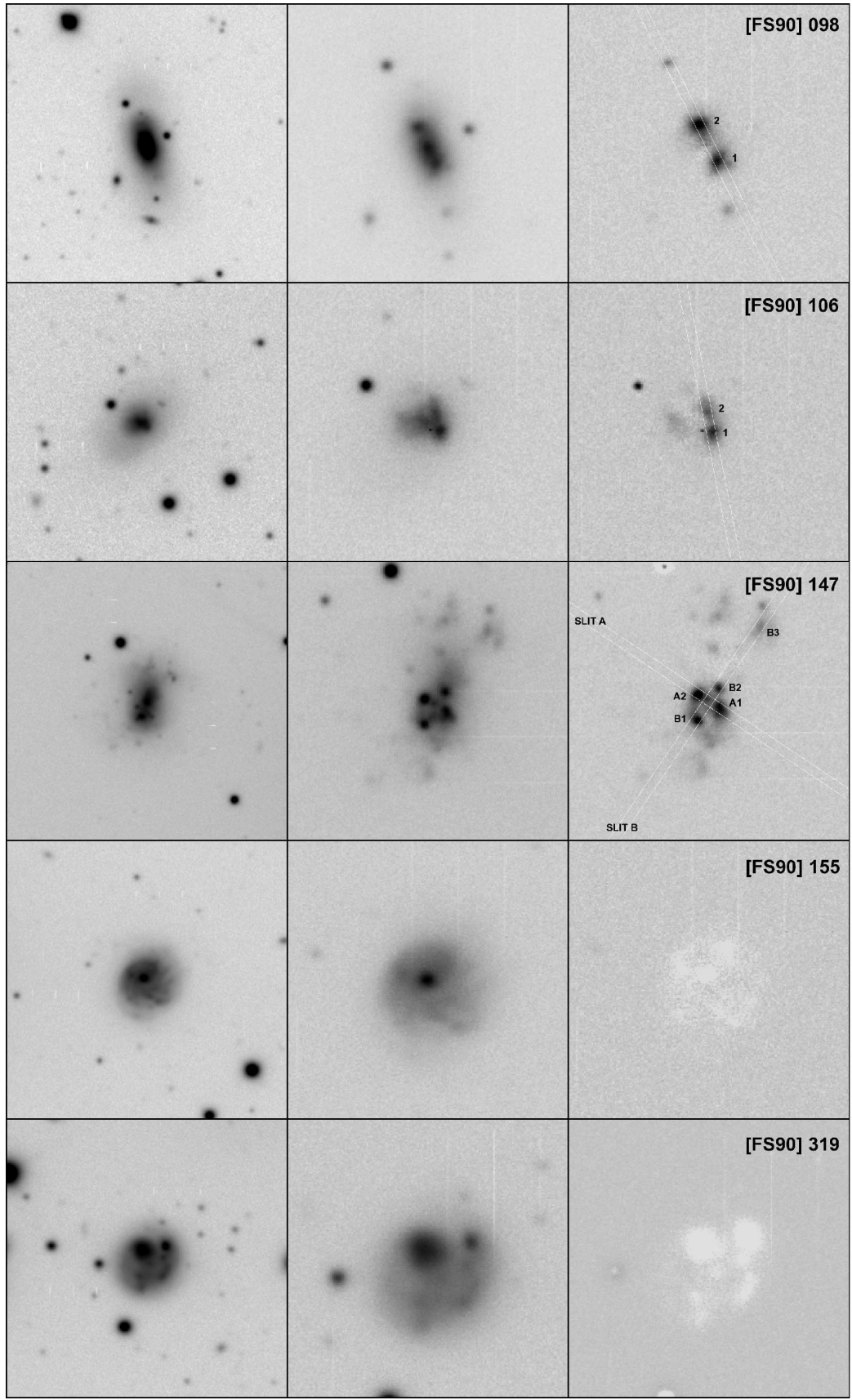}
\begin{center}
\caption{Continuum $R$ G0326 (left panel, FOV $1^\prime\times1^\prime$), 
$H\alpha$ G0336 (middle panel, FOV $0.5^\prime\times0.5^\prime$), and $H\alpha$ 
images (right panel, FOV $0.5^\prime\times0.5^\prime$) of the Antlia dwarf star-forming 
candidates observed with GMOS at Gemini South. 
The $1^{\prime\prime}$ GMOS slit and the two slits A and B used for FS90-147 are drawn 
in white in the right column plots, including most of the net $H\alpha$ emission knots. 
} 
\label{fig1}
\end{center}
\end{figure}

\begin{figure}[p]
\centering
\includegraphics[angle=0,width=9.5cm]{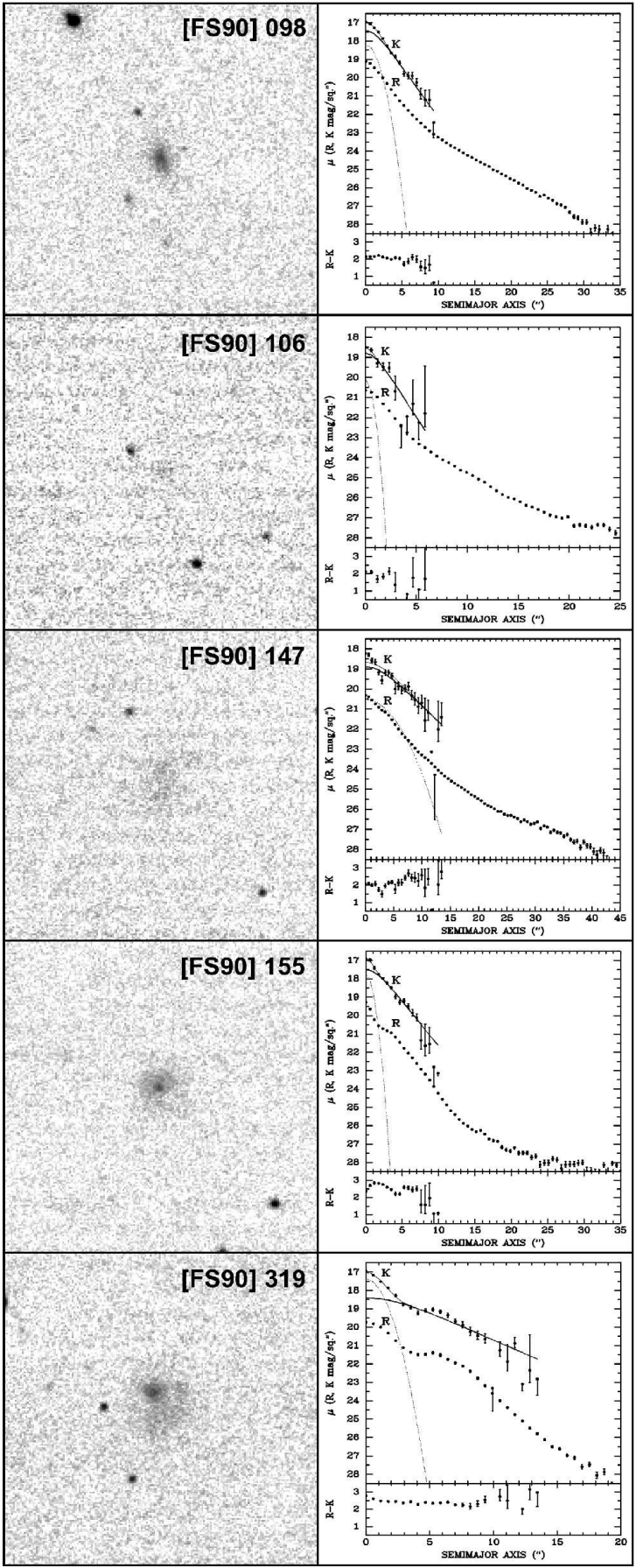}
\begin{center}
\caption{VISTA VHS $K_S$ images (left panel, FOV $1^\prime \times 1^\prime$, normal orientation)
and surface brightness profiles in $K_S$ and $R$ of the five Antlia BCD candidates. For the $K_S$ 
fits, we plot the sech component (continuous thick line), the Gaussian component (dashed line), and 
the sum of the two (continuous thin line). The colour profiles $R - K_S$ are plotted at the bottom 
of each graph. } 
\label{fig2}
\end{center}
\end{figure}

\begin{figure}[p]
\centering
\includegraphics[angle=0,width=8cm]{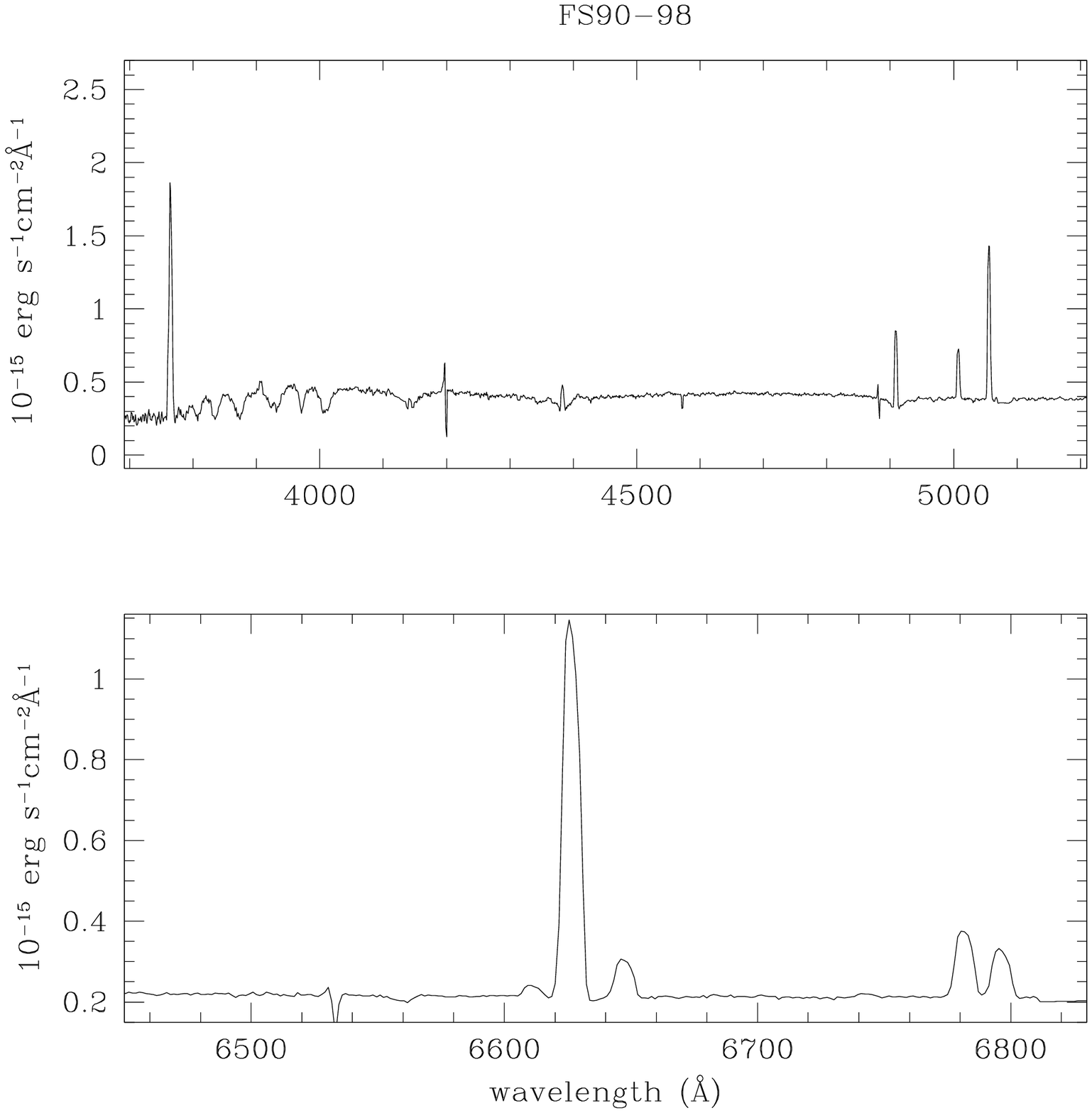}
\includegraphics[angle=0,width=8cm]{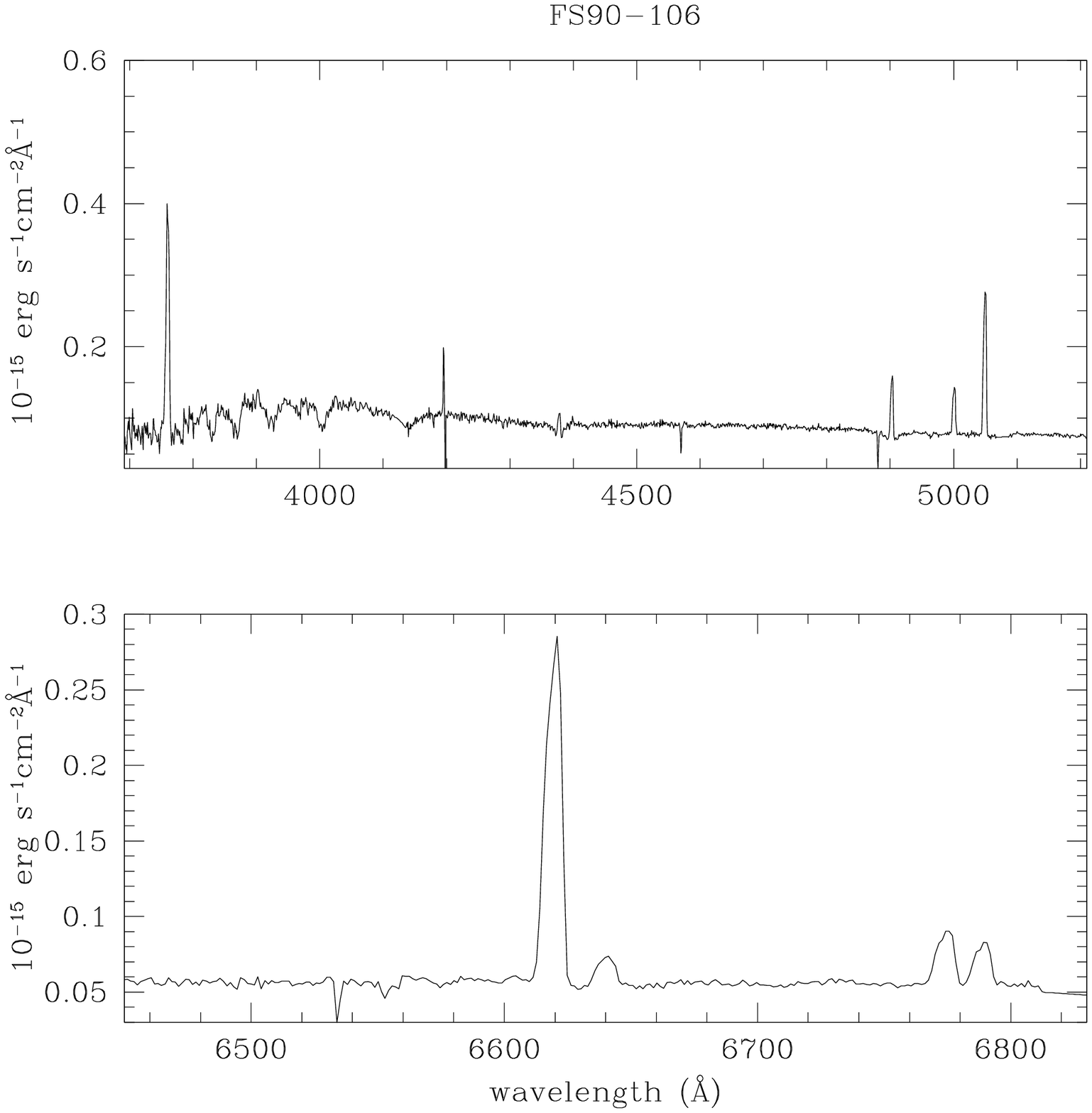}
\begin{center}
\caption{ 
Blue and red spectra (shown above and below), in units of 10$^{-15}$ erg s$^{-1}$ cm$^{-2}$ \AA$^{-1}$, 
of the star-forming candidates in our Antlia cluster: FS90-98 (integrating knots 1 and 2), FS90-106 (knot 1), 
FS90-147A (integrated knots A1 and A2), and FS90-147B (integrated knots B1 and B2). }
\label{fig3}
\end{center}
\end{figure}

This section discusses the oxygen abundances derived using the strong-line 
method by Marino et al (\cite{mar13}). 
Both emission knots of FS90-098 in the $H\alpha$ image (labelled 1 and 2 
in the upper right panel of the Figure~\ref{fig1}) were spectroscopically resolved. 
In Table~\ref{table3} we include their integrated line fluxes (column 2) and their 
individual measurements (columns 3 and 4). 
The oxygen abundances estimated for the two knots and the integrated spectrum (i.e., 
knot 1 + knot 2), $12+\log(O/H) = 8.28$ dex, are consistent within the the error of 
the Marino et al. (\cite{mar13}) calibration (0.2 dex), which is also consistent 
with the BCD classification for this galaxy. 

The two visible knots in the $H\alpha$ image of FS90-106 (labelled 1 and 2 in 
the right-hand panel of Figure~\ref{fig1}) could not be disentangled spectroscopically with 
GMOS. We measured the lines from the integrated spectrum, corresponding to the sum of the 
emission from knots 1 and 2 (see column 5 of Table 3), and found  $12+\log(O/H) = 8.25$ dex, 
which is also consistent with the BCD classification for this galaxy. 

For FS90-147, four bright knots were resolved with GMOS in the two slit configurations. 
From the two bright knots aligned along slit A (labelled A1 and A2 in the middle right 
panel of Figure~\ref{fig1}), only A1 could be spectroscopically resolved, and it shows 
$12+\log(O/H) = 8.17$ dex. 
For the other pair aligned along slit B (knots B1 and B2), we derived very similar 
oxygen abundances: 12 + log(O/H) = 8.20 for knot B1 and $12+\log(O/H) = 8.21$ dex for knot B2. 
We could also extract the spectrum of a fainter emission knot within slit B (marked as B3) 
for which we obtained $12+\log(O/H) = 8.20$ dex. Considering the typical uncertainty of 
0.2 dex from the calibration used in this work, we fouud no evidence of a significant 
metallicity variation across the galaxy FS90-147B. The abundance of this galaxy, averaged 
to $12+\log(O/H) = 8.20$ dex based on the measurements for the individual knots, that 
agrees with the metallicity values found for most of BCDs.

\subsection{Fundamental relations}

We study the classic two-dimensional correlations based on the physical and chemical data 
of our small star-forming sample in Antlia, with the aim of comparing to past similar samples 
in the Local Volume, Virgo, Fornax, and Hydra clusters (Vaduvescu et al. 
\cite{vad05,vad06,vad07,vad08,vad11}). 

To calculate absolute magnitudes, we used the following cluster distance modulus: 
$DM=32.73$ mag for Antlia (Dirsch et al. \cite{dir03}), 
$DM=31.51$ mag for Fornax (Blakeslee et al. \cite{bla09}), 
$DM=33.31$ mag for Hydra (Jensen at al. \cite{jen99}), and 
$DM=30.62$ mag for most galaxies in Virgo (Freedman et al. \cite{fre01} - based on 
Cepheids), except VCC~24 and VCC~144, for which we adopt $DM=32.53$ mag from Gavazzi et al 
(\cite{gav05}), who found them to be members of the most distant Virgo clouds W and M. 
For the LV objects, we adopt literature distances derived according to best estimators 
(mostly Cepheids and TRGB methods), all quoted with references in our past papers. For each 
object, we use the galactic extinction in $K_S$ taken from NED, although this involved very 
small corrections. 

Figure~\ref{fig4} plots the central surface brightness $\mu_{0SK}$ (in magnitudes per
square arcsec) versus the absolute magnitude $M_{SK}$, both measured based on the $K_S$ 
sech fit, which is known to quantify the old extended light component of the star-forming 
galaxies better. The well known linear correlation holds for all five Antlia galaxies (plotted 
with crosses and labelled in magenta), towards the bright luminosity (BCD) end. 
Figure~\ref{fig5} shows the scale radius $r_{0SK}$ (derived from the sech fit and measured 
in kpc) versus the sech absolute magnitude $M_{SK}$. The plot shows a trend, but the 
spread is larger for cluster objects compared with the LV, with the Antlia objects 
spreading around the composite sample. 
Figure~\ref{fig6} plots the isophotal radius $r_{22K}$ (based on the sech model and 
measured in kpc) versus the sech absolute magnitude $M_{SK}$. The known relation 
holds for both LV and cluster objects including Antlia, with larger galaxies being brighter. 
Figure~\ref{fig7} presents the oxygen abundance $12+log(O/H)$ versus the sech absolute 
magnitude $M_{SK}$. The known trend holds for both LV and cluster objects including 
Antlia, testing whether brighter objects are more chemically evolved. 

Unfortunately, no galaxy was observed in radio yet for deriving HI line widths $W_{20}$, 
and also the 21cm fluxes are unknown for the entire sample, thus we cannot check the dwarf 
fundamental plane (FP, Vaduvescu et al. \cite{vad08}; McCall et al. \cite{mcc12}), neither the 
mass - metallicity relation, nor the closed box evolution (Vaduvescu, McCall, and Richer, 
\cite{vad07}). 


\section{Conclusions}
\label{conclusions}

Five star-forming dwarf galaxy candidates were selected taking from the scarce literature 
in the Antlia cluster with the aim of continuing our studies of physical and chemical properties 
of BCDs and dIs in the Local Volume and other nearby clusters (Virgo, Fornax, and Hydra). 
Deep $H\alpha$ and $R$ imaging was obtained for these targets in arcsec seeing Band 3 
conditions using Gemini South with the GMOS-S camera. 

Three galaxies (FS90-98, FS90-106, FS90-147) out of five show $H\alpha$ emission and a few 
knots. From our GMOS spectroscopic observations, we confirm that these three galaxies are 
members of the Antlia cluster. We derive the oxygen abundance for these objects and find 
values that fall in the metallicity range derived for most of the BCDs. 
Shallower archival VHS $K_S$ images for all five targets were reduced and studied further, 
showing $K_S$ surface-brightness profile fits consistent with BCD 
classification, namely a sech law to account for the extended component plus a Gaussian 
to fit the inner outburst. 

FS90-155 and FS90-319 could not be confirmed as star-forming cluster candidates, and their 
membership status remains uncertain. FS90-155 shows some flocculent structure in both $R$ and 
$K_S$ images resembling spiral arm structures and showing no $H\alpha$ emission. 
FS90-319 presents a flat flux distribution in its inner region that has one non-central diffuse 
emission visible in both visible and NIR and no $H\alpha$ emission, so their membership 
status remains uncertain. 

Two-dimensional physical and chemical relations were studied using $K_S$ sech data for the 
whole Antlia sample to probe wheather known relations (specifically the size-luminosity and 
luminosity-metallicity) are followed by dwarfs in the LV and the Virgo, Fornax and Hydra 
clusters. Future studies could benefit from deeper NIR imaging and radio data in the aim to 
compare star-forming dwarf formation and galaxy evolution in the nearby Universe.


\begin{acknowledgements}
OV thanks to the Gemini TACs for the time allocation at Gemini South Observatory 
(programmes GS-2010A-Q-51 and GS-2012A-Q-59) which is operated by the Association of Universities 
for Research in Astronomy, Inc., under a cooperative agreement with the NSF on behalf of the Gemini 
partnership: the National Science Foundation (United States), the Science and Technology Facilities 
Council (United Kingdom), the National Research Council (Canada), CONICYT (Chile), the Australian 
Research Council (Australia), Ministerio da Ciencia e Tecnologia (Brazil) and Ministerio de Ciencia, 
Tecnologia e Innovacion Productiva  (Argentina). 
VISTA VHS raw data was obtained from the ESO Science Archive Facility under request number 72524. 
CK has been funded by the project AYA2010-21887-C04-01 from the Spanish PNAYA. 
LPB, AVSC and JPC acknowledge financial support from Consejo Nacional de Investigaciones 
Cient\'ificas y T\'ecnicas de la Rep\'ublica Argentina, Agencia Nacional de Promoci\'on Cient\'ifica 
y Tecnol\'ogica (PICT 2010-0410), and Universidad Nacional de La Plata (Argentina). 
This research has made use of the NASA/IPAC Extragalactic Database (NED) which is operated by the Jet 
Propulsion Laboratory, California Institute of Technology, under contract with the National Aeronautics 
and Space Administration. Our work used IRAF, a software package distributed by the National Optical 
Astronomy Observatory, which is operated by the Association of Universities for Research in Astronomy 
(AURA) under cooperative agreement with the National Science Foundation. This research has made use of 
SAOImage DS9, developed by Smithsonian Astrophysical Observatory. 
Acknowledgements are due to Prof. Marshall L. McCall (York University, Canada) and Prof. Tom Richtler 
(University de Concepci\'on, Chile) who took part in our Gemini proposals and contributed to the sample 
selection. We thank the anonymous referee whose feedback helped us to improve the paper. 
\end{acknowledgements}


\clearpage

\begin{table}[p]
\begin{center}
\caption{The observing log of the star-forming dwarf galaxies in the Antlia cluster observed 
with Gemini South and VISTA. 
$H\alpha$ G0336) and continuum $R$ G0326 filters refer to Gemini pre-imaging, while B600/480 
and R400/780 gratings refer to blue and red spectroscopy. $K\_S$ filter refers to NIR imaging taken 
by VISTA (VHS survey mode). } 
\label{obs_log}
\begin{scriptsize}
\begin{tabular}{lrrrrr}
\hline
\hline
\noalign{\smallskip}
Galaxy & $\alpha$ (J2000) & $\delta$ (J2000) & Date (UT) &  Filter/Gratings  & Exp (s) \\
(1)    & (2)              & (3)              & (4)       &  (5)              & (6)     \\
\hline
\noalign{\smallskip}

 FS90-98    &  10:28:34.0  &  -35:27:39  & Feb 08, 2010 & $H\alpha$ G0337  &  540 \\ 
 ...        &     ...      &     ...     & Feb 08, 2010 & $r$       G0326  &  270 \\ 
 ...        &     ...      &     ...     & Apr 16, 2012 &      B600/G5323  & 4800 \\ 
 ...        &     ...      &     ...     & Apr 18, 2012 &      R400/G5325  & 2400 \\ 
 ...        &     ...      &     ...     & Feb 05, 2013 & $K_S$            &   40 \\

 FS90-106   &  10:28:50.4  &  -35:09:36  & Feb 08, 2010 & $H\alpha$ G0337  &  540 \\ 
 ...        &     ...      &     ...     & Feb 08, 2010 & $r$       G0326  &  270 \\ 
 ...        &     ...      &     ...     & Mar 25, 2012 &      B600/G5323  & 4800 \\ 
 ...        &     ...      &     ...     & Mar 25, 2012 &      R400/G5325  & 2400 \\ 
 ...        &     ...      &     ...     & Feb 05, 2013 & $K_S$            &   40 \\

 FS90-147   &  10:29:26.6  &  -35:00:53  & Feb 08, 2010 & $H\alpha$ G0337  &  540 \\ 
 ...        &     ...      &     ...     & Feb 08, 2010 & $r$       G0326  &  270 \\ 
 slit A     &     ...      &     ...     & Mar 20, 2012 &      B600/G5323  & 4800 \\ 
 slit A     &     ...      &     ...     & Mar 20, 2012 &      R400/G5325  & 2400 \\ 
 slit B     &     ...      &     ...     & Mar 25, 2012 &      B600/G5323  & 4800 \\ 
 slit B     &     ...      &     ...     & Mar 25, 2012 &      R400/G5325  & 2400 \\ 
 ...        &     ...      &     ...     & Feb 05, 2013 & $K_S$            &   40 \\

 FS90-155   &  10:29:34.8  &  -35:02:42  & Feb 08, 2010 & $H\alpha$ G0337  &  540 \\ 
 ...        &     ...      &     ...     & Feb 08, 2010 & $r$       G0326  &  270 \\ 
 ...        &     ...      &     ...     & Feb 05, 2013 & $K_S$            &   40 \\

 FS90-319   &  10:32:07.3  &  -35:02:24  & Feb 08, 2010 & $H\alpha$ G0337  &  540 \\ 
 ...        &     ...      &     ...     & Feb 08, 2010 & $r$       G0326  &  270 \\ 
 ...        &     ...      &     ...     & Feb 05, 2013 & $K_S$            &   40 \\

\noalign{\smallskip}
\hline
\hline
\end{tabular}
\end{scriptsize}
\label{table1}
\end{center}
\end{table}


\begin{table}[!t]
\begin{center}
\caption{Physical parameters of the galaxies observed in $K_S$ and $R$ bands in Antlia cluster: (1) galaxy name; (2) ellipticity; 
(3) position angle (degrees, anti-clockwise from N to E); (4) total apparent magnitude in $K_S$; (5) sech magnitude in $K_S$; 
(6) sech central surface brightness in $K_S$ (mag/arcsec$^2$); (7) sech scale radius in $K_S$ (arcsec); (8) radius (arcsec) at $K_S=22$ 
mag/arcsec$^2$; (9) total apparent magnitude in $R$; (10) $R-K_S$ total colour; (11) absolute sech $K_S$ magnitude assuming 
$(m - M) = 32.73 \pm 0.25$ for Antlia (Dirsch et al. \cite{dir03}). 
} 
\begin{minipage}{15.5cm}
\begin{tabular}{lrrrrrrrrrr}
\hline
\hline
\noalign{\smallskip}
$Galaxy$   & $e$   &  $PA$ & $m_{TK}$ &  $m_{SK}$  & $\mu_{OK}$ & $r_{OK}$ & $r_{22K}$ & $m_{TR}$ &  $R-K_S$ &  $M_{SK}$ \\
(1)        & (2)   &  (3)  & (4)      &  (5)       &  (6)       &  (7)     &  (8)      &  (9)     &   (10)   &  (11)     \\
\hline
\noalign{\smallskip}

 FS90-98   &  0.6  &  +10  &   14.00  &     14.05  &    17.43   &    2.0   &     9.7   &   16.02  &    2.02  &   -18.68  \\
 FS90-106  &  0.3  &  -50  &   15.90  &     15.82  &    18.79   &    1.4   &     5.0   &   17.02  &    1.12  &   -16.91  \\
 FS90-147  &  0.5  &  -10  &   13.84  &     13.77  &    18.89   &    4.0   &    14.3   &   16.09  &    2.25  &   -18.96  \\
 FS90-155  &  0.1  &  -20  &   13.26  &     13.24  &    17.51   &    2.2   &    10.7   &   15.69  &    2.43  &   -19.49\footnote{Hypothetical absolute magnitude assuming cluster membership (not confirmed)} \\
 FS90-319  &  0.1  &  -20  &   13.16  &     13.10  &    18.41   &    3.6   &    14.3   &   15.54  &    2.38  &   -19.63\footnote{idem} \\

\noalign{\smallskip}
\hline
\hline
\end{tabular}
\end{minipage}
\label{table2}
\end{center}
\end{table}


\clearpage

\begin{table}
\caption{Reddening-corrected line fluxes relative to F(\mbox{H$\beta$})=1 for the sample of Antlia galaxies. Please refer to the knots and slits labelled in the right-hand panel of Figure~\ref{fig1}. }

\tiny{
\begin{minipage}{15.5cm}
\label{table3}
\centering
\begin{tabular}{lccccccccc}
\hline\hline
\\
Wavelength        &    FS90-98     &   FS90-98        &   FS90-98       &   FS90-106      &  FS90-147 A     &   FS90-147 B    &   FS90-147 B    &   FS90-147 B    &   FS90-147 B \\ 
                  &   knots 1+2    &    knot 1        &    knot 2       &   knots 1+2     &    knot A1      &  knots B1+B2    &     knot B1     &     knot B2     &    knot B3   \\
    (1)           &      (2)       &     (3)          &      (4)        &      (5)        &      (6)        &       (7)       &      (8)        &      (9)        &      (10)    \\
\\
\hline
\\
3727 [O~II]       & 5.49 $\pm$ 0.23 & 4.11 $\pm$ 0.17 & 2.92 $\pm$ 0.13 & 4.82 $\pm$ 0.27 & 2.57 $\pm$ 0.13 & 3.26 $\pm$ 0.09 & 3.81 $\pm$ 0.22 & 6.09 $\pm$ 0.25 &  5.34 $\pm$ 0.72 \\
3868 [Ne~III]     & ---             & ---             & 0.19 $\pm$ 0.03 & ---             & 0.36 $\pm$ 0.06 & 0.46 $\pm$ 0.04 & 0.58 $\pm$ 0.05 & 0.72 $\pm$ 0.10 &  ---             \\
4340 H$\gamma$    & 0.48 $\pm$ 0.03 & 0.51 $\pm$ 0.04 & 0.47 $\pm$ 0.02 & 0.42 $\pm$ 0.05 & 0.43 $\pm$ 0.06 & 0.43 $\pm$ 0.05 & 0.50 $\pm$ 0.04 & 0.52 $\pm$ 0.05 &  0.53 $\pm$ 0.07 \\
4860 H$\beta$     & 1.00 $\pm$ 0.02 &1.00  $\pm$ 0.02 & 1.00 $\pm$ 0.01 & 1.00 $\pm$ 0.03 & 1.00 $\pm$ 0.02 & 1.00 $\pm$ 0.01 & 1.00 $\pm$ 0.01 & 1.00 $\pm$ 0.02 &  1.00 $\pm$ 0.03 \\   
4959 [O~III]      & 0.58 $\pm$ 0.02 & 0.55 $\pm$ 0.01 & 0.55 $\pm$ 0.01 & 0.70 $\pm$ 0.02 & 1.02 $\pm$ 0.02 & 0.93 $\pm$ 0.01 & 1.32 $\pm$ 0.02 & 1.01 $\pm$ 0.02 &  0.82 $\pm$ 0.03 \\  
5007 [O~III]      & 1.76 $\pm$ 0.03 & 1.61 $\pm$ 0.03 & 1.64 $\pm$ 0.02 & 2.10 $\pm$ 0.04 & 3.05 $\pm$ 0.04 & 2.77 $\pm$ 0.03 & 3.91 $\pm$ 0.04 & 3.06 $\pm$ 0.05 &  2.33 $\pm$ 0.07 \\ 
6563 H$\alpha$    & 2.88 $\pm$ 0.06 & 2.89 $\pm$ 0.06 & 2.85 $\pm$ 0.03 & 2.88 $\pm$ 0.08 & 2.82 $\pm$ 0.05 & 2.89 $\pm$ 0.03 & 2.91 $\pm$ 0.06 & 2.90 $\pm$ 0.06 &  2.96 $\pm$ 0.21 \\
6584 [N~II]       & 0.28 $\pm$ 0.01 & 0.31 $\pm$ 0.01 & 0.231$\pm$0.004 & 0.24 $\pm$ 0.02 & 0.16 $\pm$ 0.01 & 0.20 $\pm$ 0.01 & 0.20 $\pm$ 0.01 & 0.19 $\pm$ 0.01 &  0.19 $\pm$ 0.02 \\ 
6678 HeI          & 0.022$\pm$0.003 & ---             & 0.031$\pm$0.002 & ---             & ---             & ---             & 0.02 $\pm$ 0.01 & ---             &  ---             \\
6717 [S~II]       & 0.52 $\pm$ 0.01 & 0.56 $\pm$ 0.01 & 0.36 $\pm$ 0.01 & 0.47 $\pm$ 0.02 & 0.31 $\pm$ 0.01 & 0.36 $\pm$ 0.01 & 0.26 $\pm$ 0.01 & 0.34 $\pm$ 0.01 &  0.41 $\pm$ 0.04 \\  
6731 [S~II]       & 0.37 $\pm$ 0.01 & 0.40 $\pm$ 0.01 & 0.257$\pm$0.004 & 0.34 $\pm$ 0.01 & 0.22 $\pm$ 0.01 & 0.247$\pm$ 0.004& 0.18 $\pm$ 0.01 & 0.22 $\pm$ 0.01 &  0.28 $\pm$ 0.03 \\ 
\\
\hline
\\
$C_{H\beta}$      & 0.29 $\pm$ 0.02 & 0.21 $\pm$ 0.02 & 0.00 $\pm$ 0.01 & 0.28 $\pm$ 0.03 & 0.08 $\pm$ 0.02 & 0.21 $\pm$ 0.01 & 0.44 $\pm$ 0.02 & 0.45 $\pm$ 0.02 &  0.40 $\pm$ 0.10 \\
$n_{e} (cm^{-3})$ &  $<$ 100        & $<$ 100         & $<$ 100         &  $<$ 100        &  $<$ 100        & $<$ 100         & $<$ 100         &  $<$ 100        &  $<$ 100         \\
12+log(O/H)$_{N2}$\footnote{O/H derived from  Marino et al. (\cite{mar13}) -  N2 method} 
                  &      8.28       &      8.29       &      8.24       &      8.25       &      8.17       &      8.21       &      8.20       &      8.21       &      8.20        \\ 
\hline       
\end{tabular}
\end{minipage}
\label{table3}}
\end{table}



\clearpage

\clearpage

\clearpage

\begin{figure}[p]
\centering
\includegraphics[angle=0,width=8cm]{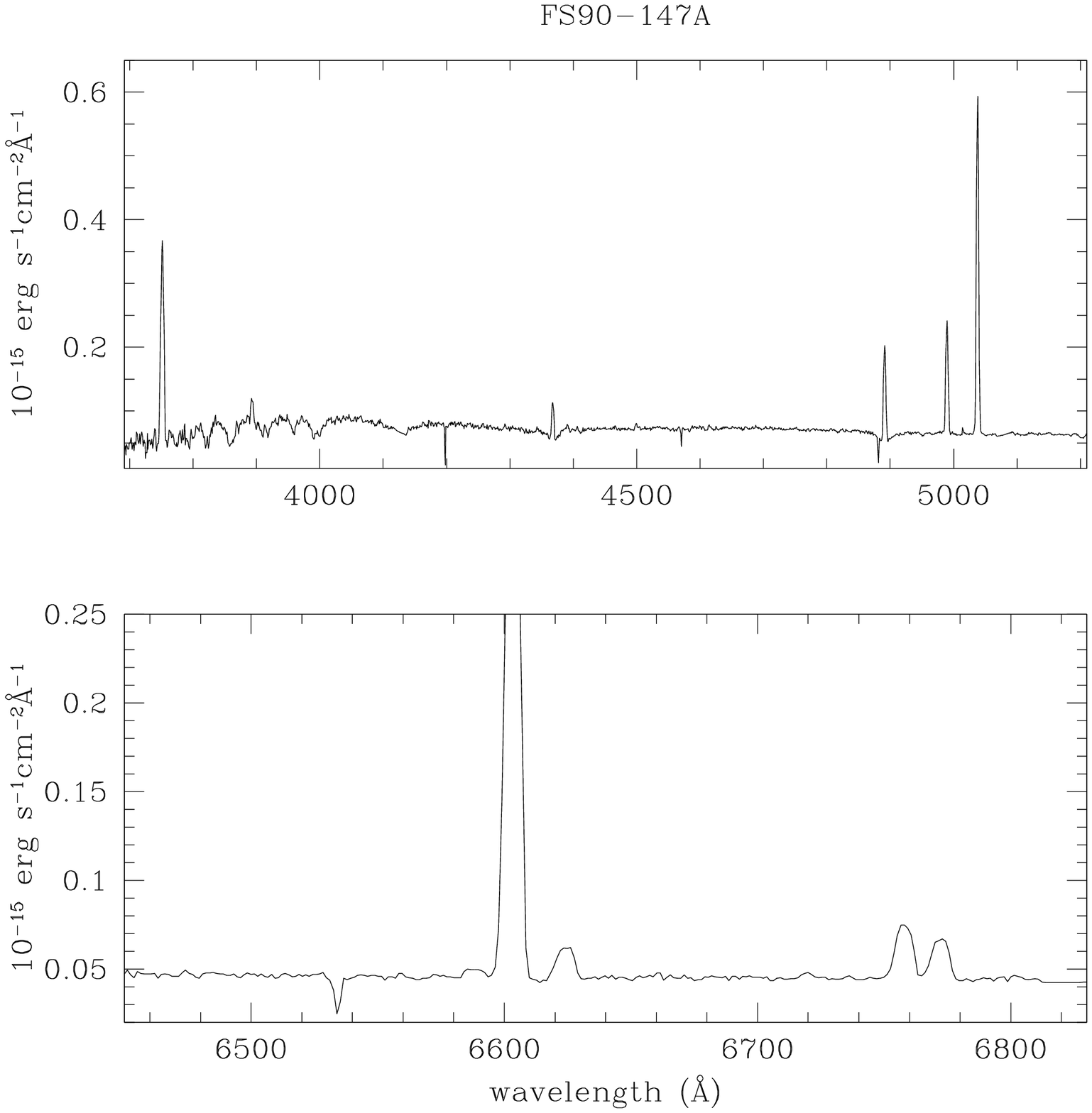}
\includegraphics[angle=0,width=8cm]{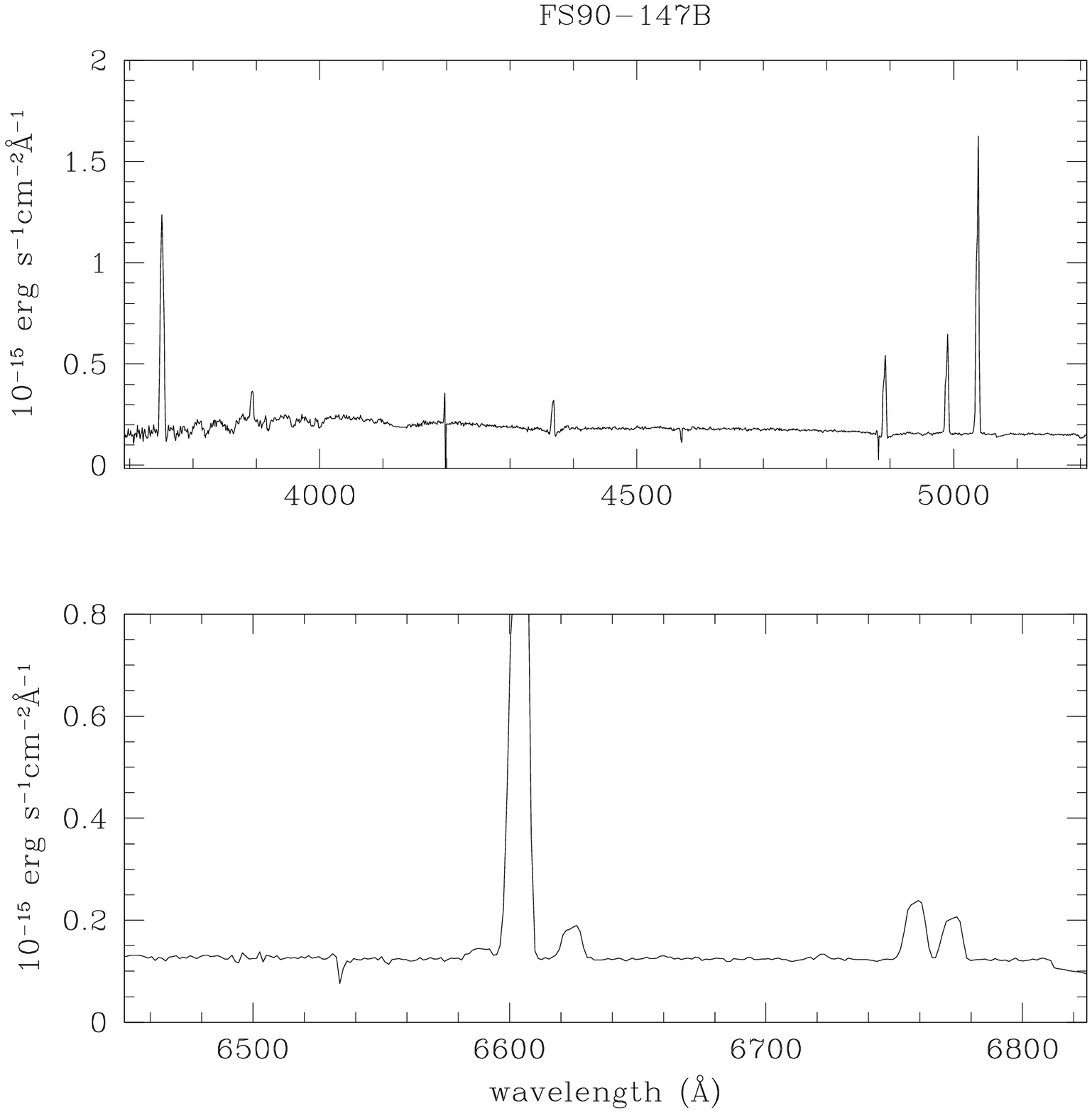}
\begin{center}
Figure 3 (continued) 
\end{center}
\end{figure}

\begin{figure}[p]
\centering
\includegraphics[angle=0,width=8cm]{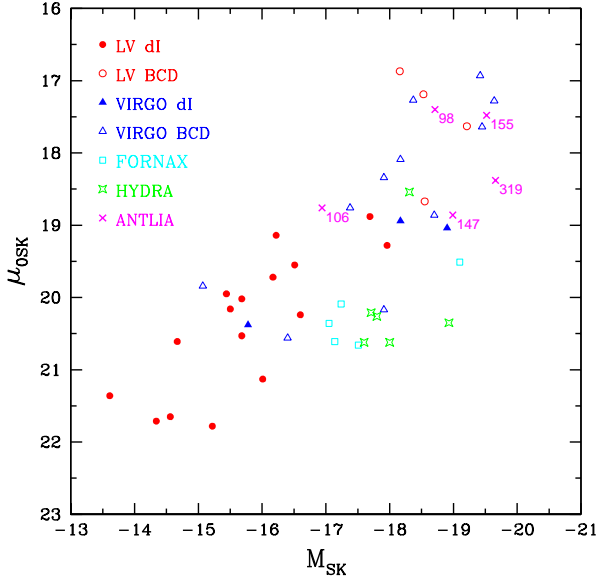}
\begin{center}
\caption{ 
Sech central surface brightness $\mu_{0SK}$ (mag $arcsec^{-2}$) versus sech absolute magnitude $M_{SK}$ for the star-forming 
dwarfs in the Local Volume, Virgo, Fornax, Hydra, and Antlia samples. The well known linear trend is probed by the Antlia galaxies. }
\label{fig4}
\end{center}
\end{figure}

\begin{figure}[p]
\centering
\includegraphics[angle=0,width=8cm]{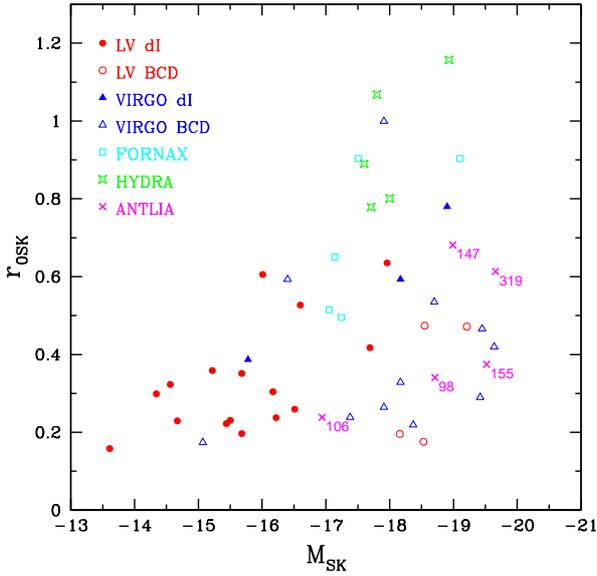}
\begin{center}
\caption{ 
Sech scale radius $r_{0SK}$ (measured in kpc) versus sech absolute magnitude $M_{SK}$ for the star-forming 
dwarfs in the Local Volume, Virgo, Fornax, Hydra, and Antlia samples. The distribution is quite widespread and includes 
the Antlia galaxies. }
\label{fig5}
\end{center}
\end{figure}

\begin{figure}[p]
\centering
\includegraphics[angle=0,width=8cm]{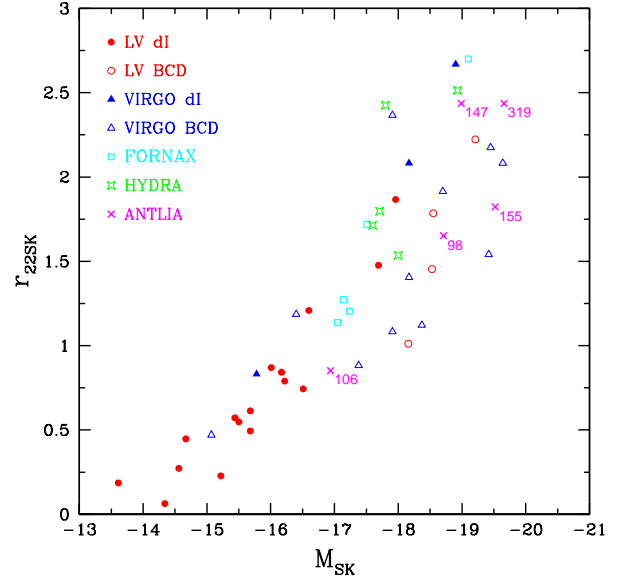}
\begin{center}
\caption{ 
Sech isophotal semi-major axis $r_{22SK}$ (measured in kpc) versus sech absolute magnitude $M_{SK}$ for the star-forming dwarfs 
in the Local Volume, Virgo, Fornax, Hydra, and Antlia samples. The well known trend is probed by the Antlia galaxies. }
\label{fig6}
\end{center}
\end{figure}

\begin{figure}[p]
\centering
\includegraphics[angle=0,width=8cm]{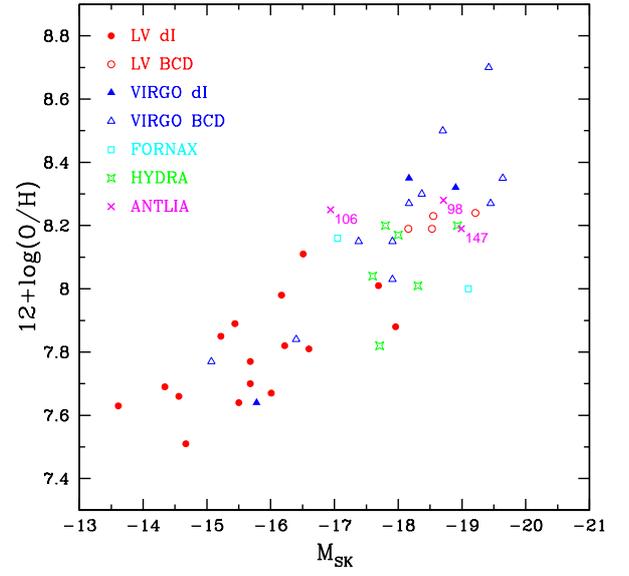}
\begin{center}
\caption{ 
Oxygen abundance $12+log(O/H)$ versus sech absolute magnitude $M_{SK}$ for the star-forming dwarfs in the Local Volume, 
Virgo, Fornax, Hydra, and Antlia samples. The well known linear trend is probed by three Antlia galaxies with available spectra. }
\label{fig7}
\end{center}
\end{figure}

\end{document}